\documentclass{article}
\usepackage{ijcai20}

\usepackage{times}

\usepackage{soul}
\usepackage{url}
\usepackage[hidelinks]{hyperref}
\usepackage[utf8]{inputenc}
\usepackage[small]{caption}
\usepackage{graphicx}
\usepackage{amsmath}
\usepackage{booktabs}
\usepackage{enumerate}
\usepackage{subfigure}
\usepackage{algorithm}
\usepackage{algorithmic}
\usepackage{color} 
\usepackage{amsmath}
\usepackage{amsfonts}
\usepackage{booktabs}
\usepackage{multirow}
\usepackage{tikz}
\newcommand*{\circled}[1]{\lower.7ex\hbox{\tikz\draw (0pt, 0pt)%
    circle (.45em) node {\makebox[1em][c]{\small #1}};}}

\title{Learning to Accelerate Heuristic Searching for Large-Scale Maximum Weighted b-Matching Problems in Online Advertising}

\author{
Xiaotian Hao$^1$\and
Junqi Jin$^{2\ *}$\and
Jianye Hao$^{1,3,4}\ $\thanks{Corresponding authors.}\and
Jin Li$^2$\and
Weixun Wang$^1$\and
Yi Ma$^1$\and
Zhenzhe Zheng$^5$\and
Han Li$^2$\and
Jian Xu$^2$\And
Kun Gai$^2$
\affiliations
$^1$College of Intelligence and Computing, Tianjin University \\
$^2$Alibaba Group, Beijing \\
$^3$Noah's Ark Lab, Huawei \\
$^4$Tianjin Key Lab of Machine Learning \\
$^5$Shanghai Jiao Tong University
\emails
\{xiaotianhao,jianye.hao,wxwang, mayi\}@tju.edu.cn, zhengzhenzhe220@gmail.com,
\{junqi.jjq,echo.lj,lihan.lh,xiyu.xj\}@alibaba-inc.com, jingshi.gk@taobao.com
}

\begin{document}

\maketitle
\begin{abstract}
Bipartite b-matching is fundamental in algorithm design, and has been widely applied into economic markets, labor markets, etc. These practical problems usually exhibit two distinct features:
large-scale and dynamic, which requires the matching algorithm to be repeatedly executed at regular intervals. However, existing exact and approximate algorithms usually fail in such settings due to either requiring intolerable running time or too much computation resource.
To address this issue,
we propose \texttt{NeuSearcher} which leverages the knowledge learned from previously instances to solve new problem instances.
Specifically, we design a multichannel graph neural network to predict the threshold of the matched edges weights, by which the search region could be significantly reduced. We further propose a parallel heuristic search algorithm to iteratively improve the solution quality until convergence.
Experiments on both open and industrial datasets demonstrate that \texttt{NeuSearcher} can speed up 2 to 3 times while achieving exactly the same matching solution compared with the state-of-the-art approximation approaches.

\end{abstract}

\section{Introduction}
Bipartite b-matching is one of the fundamental problems in computer science and operations research. Typical applications include resource allocation problems, such as job/server allocation in cloud computing and product recommendation\cite{de2011social} and advertisement (ad) allocation \cite{agrawal2018proportional} in economic markets. It has also been utilized as an algorithmic tool in a variety of domains, including document clustering \cite{dhillon2001co}, computer vision \cite{zanfir2018deep}, and as a subroutine in machine learning algorithms. The focus of this paper is on large-scale real-world bipartite b-matching problems, which usually involve billions of nodes and edges and the graph structure dynamically evolves. One concrete example is the ads allocation in targeted advertising.

In targeted advertising, a bipartite graph connects a large set of consumers and a large set of ads. We associate a relevance score (e.g., click through rate) to each potential edge of a consumer to an ad, which measures the degree of interest a consumer has over an ad. Each edge then can be seen as an allocation from an ad to a consumer with the corresponding score. Due to the business reasons, for each consumer and ad, there are cardinality constraints on the maximum number of edges that each vertex can be allocated.
The goal of the ad allocation is to search for a maximum weighted b-matching: selecting a subset of edges with the maximum total scores while satisfying the cardinality constraints.

The first exact algorithm for b-matching was the Blossom algorithm \cite{edmonds1965maximum}. After that, several exact b-matching approaches have been proposed, such as branch and cut approach \cite{padberg1982odd}, cutting plane technique \cite{grotschel1985solving} and belief propagation \cite{bayati2011belief}. Interested readers can refer to \cite{muller2000implementing} for a complete survey. The time complexity of these exact matching algorithms is proportional to the product of the numbers of edges and vertices \cite{naim2018scalable}. In advertising, there exist hundreds of millions of consumers and ads with billions of edges, which makes the exact algorithms computationally infeasible.

Another challenge in advertising is that the bipartite graph dynamically evolves with time, e.g., consumers' interests over ads may be different in different period, which changes the edges' scores. For this reason, the matching problem has to be repeatedly solved (e.g., hour-to-hour) to guarantee matching performance. This requires that an algorithm must compute the solution fast to satisfy the online requirements.
Though we can use approximate algorithms with parallel computation to reduce the new solution computation time \cite{de2011social,khan2016efficient},
all of them starts the solution computation of each new problem instance from scratch. It would be more desirable if the knowledge learned from previous solved instances can be (partially) transferred to the new ones (similar but not exactly the same) to further reduce the computation time.

For this purpose, we investigate whether we can leverage the representation capability of neural networks to transfer the knowledge learned from previous solved instances to accelerate the solution computing on similar new instances. In this paper, we propose a parallelizable and scalable learning based framework \texttt{NeuSearcher} to accelerate the solution computing for large-scale b-matching. Our contributions in this paper can be summarized as follows: (1) We propose \texttt{NeuSearcher} which integrates machine learning to transfer knowledge from previous solved instances to similar new ones, which significantly reduces the computational cost and reaches up to 2-3 times faster than the state-of-the-art approximation algorithms. (2) We build a predictive model to predict the threshold of matched edges weights to reduce the search region of the solution space. Then, we design a heuristic search algorithm to ensure the solution quality and convergence. We show that it is guaranteed that the \texttt{NeuSearcher}'s solution quality is exactly the same with the state-of-the-art approximation algorithms. (3) As the bipartite graph in advertising is unbalanced, i.e., the number of consumers is extremely larger than that of ads, we design a multichannel graph neural network (GNN) to improve the accuracy of the predictive model. (4) Experiments on open and industrial large-scale datasets demonstrate that \texttt{NeuSearcher} can compute nearly optimal solution much faster than the state-of-the-art approaches.

\section{Maximum Weighted b-Matching Problem}
\label{definition}
In a targeted advertising system, there are a set of ads $\mathbb{A}$=$\{a_1,...,a_m\}$, which are to be delivered to a set of consumers $\mathbb{C}$=$\{c_1,...,c_n\}$. For each $a_i$ and $c_j$, we measure the interest of consumer $c_j$ in ad $a_i$ with a positive weight $\text{w}(a_i, c_j)$ (e.g., click through rate). Each ad has to pay a fee to the platform once been displayed to (or clicked by) a consumer. Since the advertising budget is limited, each advertiser aims to pick out a limited number of their best audiences from $\mathbb{C}$ to deliver its ad to maximize the profits. Hence, we set a capacity constraint $b(a_i)$ on the number of consumers that each ad $a_i$ can match. Besides, to avoid each consumer $c_j$ receiving too many ads, we enforce a capacity constraint $b(c_j) $ on the number of ads that are matched to $c_j$. The goal is to find a subset of edges $M \subseteq \mathbb{E}$ such that the capacity constraints for each ad and consumer are satisfied, while maximizing the total weight of the matching. Such an edge set $M$ is referred to as a maximum weighted b-matching.
Formally, we have:
\begin{align}
\mathop{\text{max}}_{\mathcal{X}} \sum_{(a_i,c_j)\in\mathbb{E}}&x_{i,j} \text{w}(a_i, c_j)\\
\text{s.t.}~\sum_{c_j\in\mathbb{C}}&x_{i,j}\le b(a_i), \forall a_i\in\mathbb{A},\\
\sum_{a_i\in\mathbb{A}}&x_{i,j}\le b(c_j), \forall c_j\in\mathbb{C} 
\end{align} 
where $\mathcal{X}\!=\!\{x_{i,j}|(a_i,c_j)\in\mathbb{E}\}$ is the decision variable, $x_{i,j} \in\{0,1\}$ indicates whether edge $(a_i,c_j)$ is included in $M$.

However, the relationship between consumers and advertisers changes frequently in practice. The main reason is that the consumers' interests are evolving, which changes the edge weight $\text{w}(a_i, c_j)$ of the matching problem. Therefore, similar problem instances have to be repeatedly solved for better matching qualities. In the following of this paper, we consider these repeatedly solved b-matching problem instances $\mathcal{I}=\{I_1,...,I_N\}$ are generated from the same distribution $\mathbb{D}$. And we are interested in investigating whether we can leverage neural network to transfer the knowledge learned from previous solved instances to accelerate the solution computing on new instances. Though, some recent works incorporate machine learning methods to solve combinatorial optimization (CO) problems, e.g., learning to solve the Traveling Salesman Problem \cite{vinyals2015pointer,khalil2017learning,li2018combinatorial} and Mixed Integer Programming problems \cite{he2014learning,chen2019learning,ding2019optimal}, no researches aim to solve the practical large-scale b-matching problems and these existing methods are not applicable in our case. The reason is that these methods usually model the CO problem as a sequential decision-making process via imitation learning or 
reinforcement learning, whose time complexity is proportional to the edge number. The time complexity is too high. Besides, these approaches can only be applied to small problem instances, e.g., problems with thousands nodes or edges. But the problem we consider is in billion scale.

Next, we start by analyzing the core idea and the bottlenecks of the state-of-the-art parallel approximation approaches. Then, we derive which form of knowledge can be transferred from previous solved problem instances to new ones and propose our \texttt{NeuSearcher} framework.

\section{Bottleneck of Approximation Approaches}
\label{greedy-problem}

\begin{figure*}[t]
\setlength{\abovecaptionskip}{0.15cm}
\centering
\subfigure[b-matching problem example] {\includegraphics[width=1.64in,angle=0]{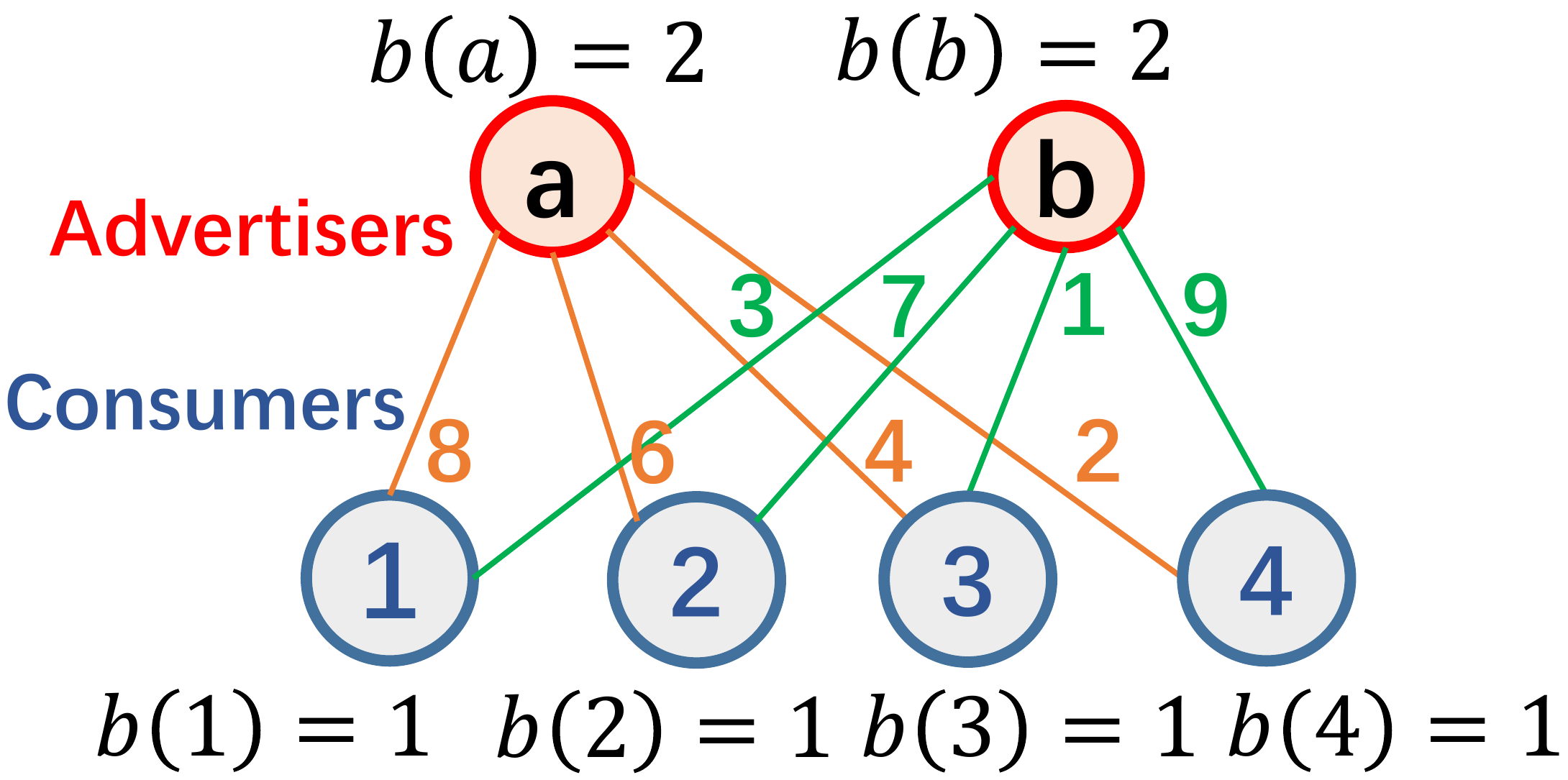}}
\subfigure[step 1: sorting neighbors] {\includegraphics[width=1.64in,angle=0]{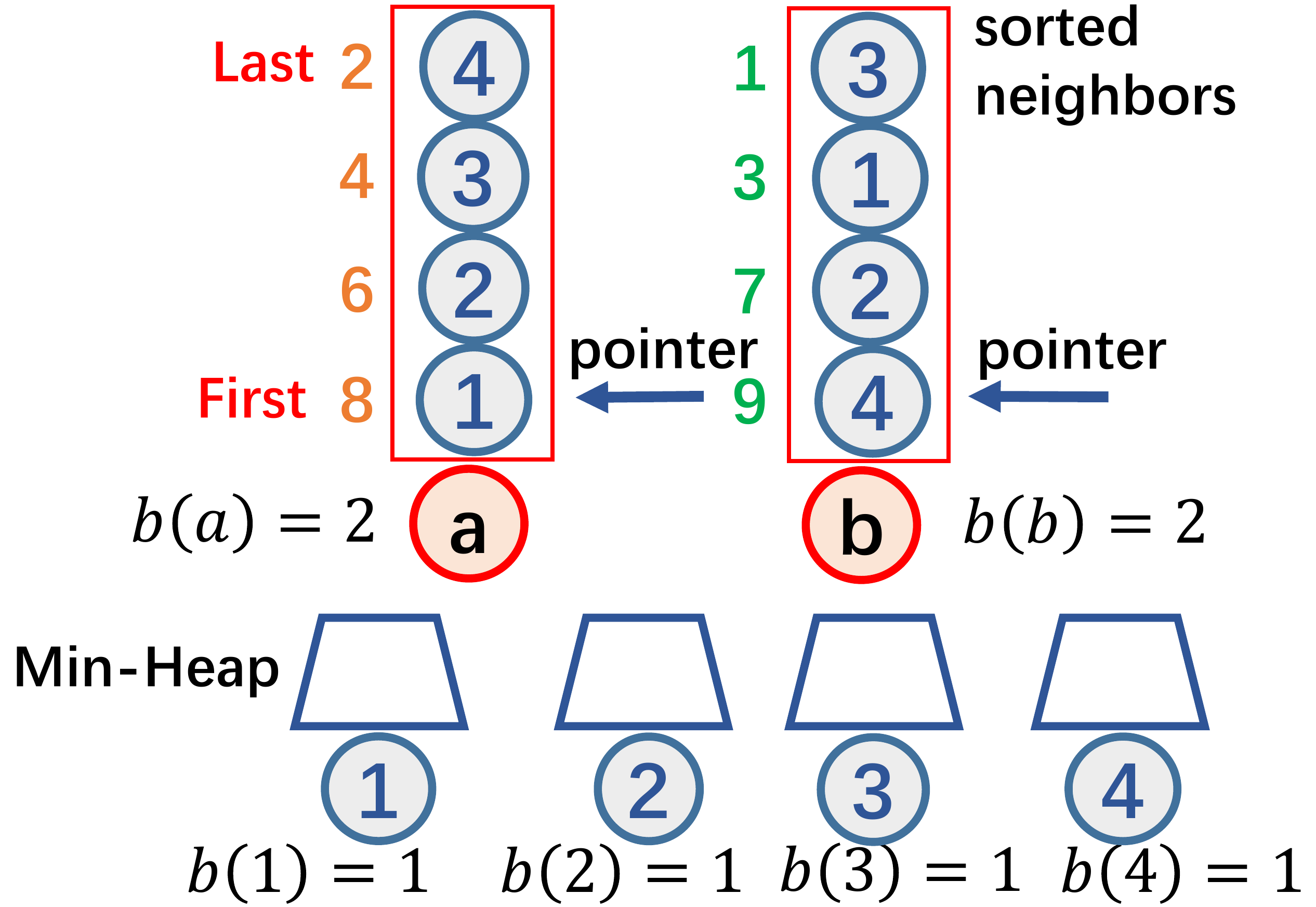}}
\subfigure[step 2: iteration \#1] {\includegraphics[width=1.64in,angle=0]{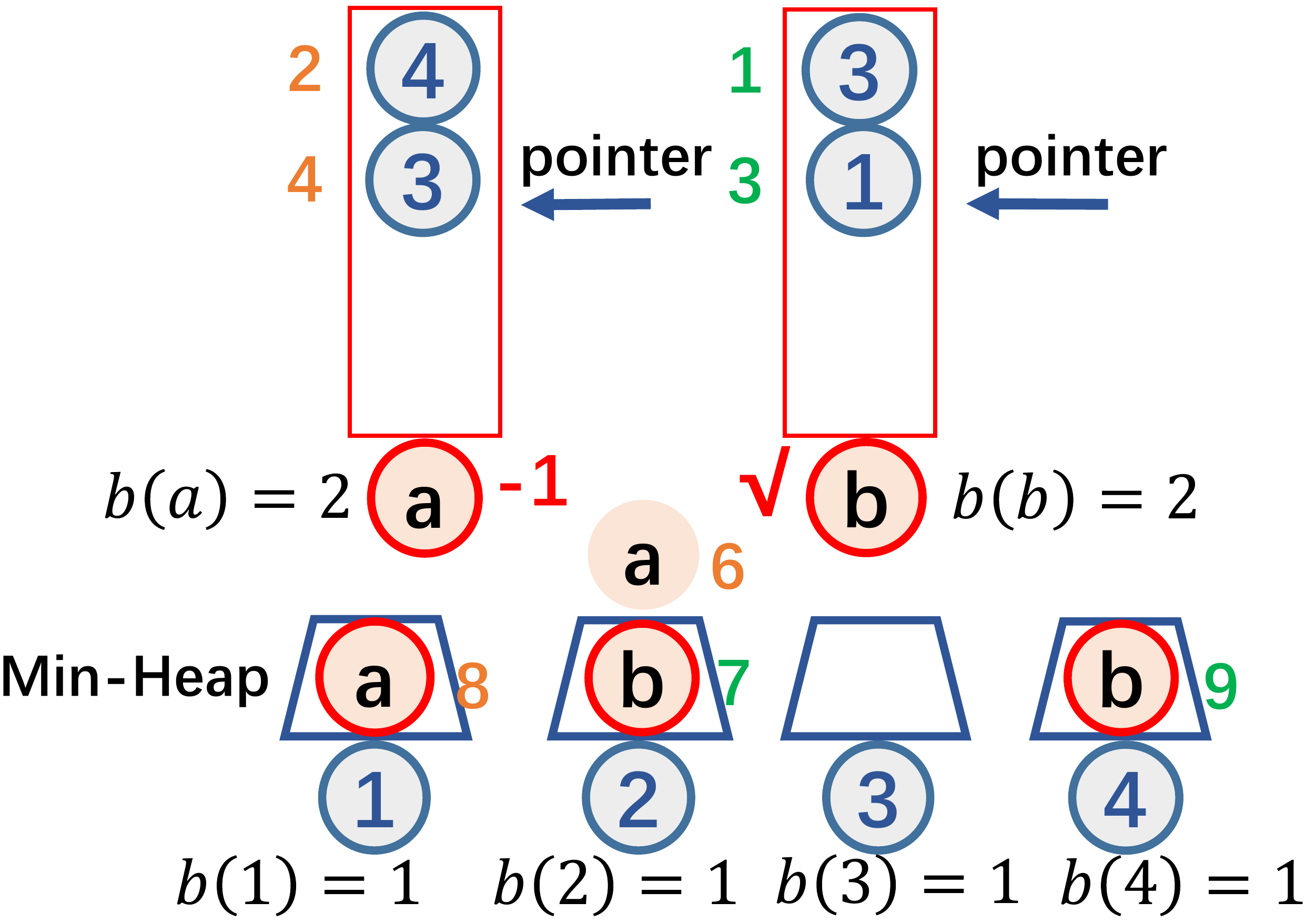}}
\subfigure[step 3: iteration \#2] {\includegraphics[width=1.64in,angle=0]{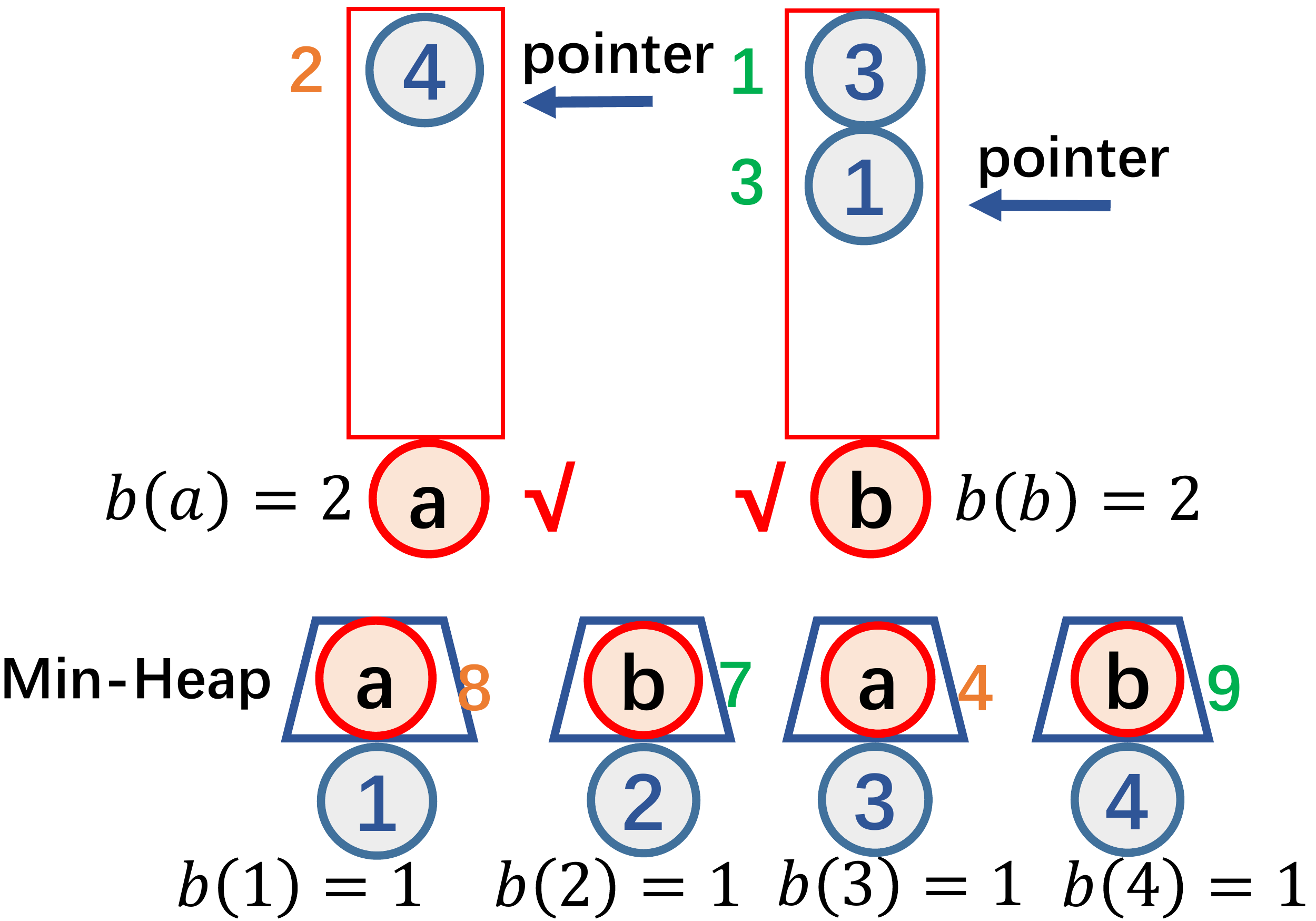}}
\caption{An illustration of the core idea of the paralleled greedy approaches (best viewed in color).}
\label{Figure:b-filter}
\end{figure*}

The greedy algorithm is the most commonly used approximation approach in practice. It works by sorting all the edges globally in descending order of their weights. After that, it picks edges one by one from the heaviest to the lightest only if the capacity constraints on both end points of an edge are satisfied. But, if the graph has billions of edges: (1) the global sorting of all edges costs too much time and becomes a bottleneck. (2) the sequential nature of adding edges to the solution is slow. Accordingly, paralleled greedy approaches are proposed, e.g., GreedyMR and b-suitor \cite{khan2016efficient,naim2018scalable}, which are the state-of-the-art parallelizable approximate methods for computing b-matching solutions. We explain the core idea of these methods through a simple example. As shown in Fig \ref{Figure:b-filter}(a), there are 2 ad vertices $a$ and $b$, both of which have a capacity constraint $b(a)$ = $b(b)$ = $2$. There are 4 consumer vertices whose indices range from $1$ to $4$, all of which have a constraint $b(1)$ =,...,= $b(4)$ = 1. And there is a weight $\text{w}(a_i,c_j)$ marked alongside each edge (i.e., 3,7,1,9 in green and 8,6,4,2 in orange). The paralleled greedy approach works iteratively as:
\begin{itemize}
\vskip -0.1cm
\setlength{\itemsep}{0.5pt}
\setlength{\parsep}{0.5pt}
\setlength{\parskip}{0.5pt}
\item At the initial step (Figure \ref{Figure:b-filter}(b)), each consumer $c$ initializes an empty minimum heap of size $b(c)$=1 (shown as blue trapezoids). The target is to reserve the top-$b(c)$ neighbors with largest edge weights for each consumer node $c$. After initialization, each ad sorts its neighbors in parallel by descending order according to their edge weights. The sorted consumer nodes are shown in the 2 red rectangles. Each ad maintains a pointer pointing to the vertex with the largest weight of the remaining sorted neighbors.
\item At the first iteration (Figure \ref{Figure:b-filter} (c)), each ad vertex $v$ pours out the first $b(v)$=$2$ vertices from the sorted neighbors and tries to put the 2 edges into the minimum heap of the corresponding consumer vertices. However, since the capacity of each minimum heap is limited ($b(c)$=1), ad vertex with the smallest edge weight will be squeezed out when the minimum heap is full. For example, in Figure \ref{Figure:b-filter} (c), vertex $a$ is squeezed out from the minimum heap of vertex $2$ because its weight is $6$, which is smaller than the competitor vertex $b$'s weight of $7$.
\item After the first iteration (Figure \ref{Figure:b-filter}(c)), because vertex $a$'s second neighbor with edge weight 6 is squeezed out, it moves its pointer to the next consumer and pours out 1 more consumer with index of $3$, whose edge weight 4 is the largest among the remained sorted neighbors.
\item After the second iteration (Figure \ref{Figure:b-filter}(d)), all ad vertices have successfully reserved two neighbors, thus the iteration stops and the solution of the b-matching is reserved in the minimum heaps of the consumer vertices.
\end{itemize}
\vskip -0.1cm
Intuitively the above process can be understood as a process of ``pouring water". Each ad behaves like a ``kettle" and each consumer behaves as a ``priority-cup'' (ads with smaller weights are easier to get out of the cup). Each ad iteratively pours out the sorted neighbors until the accepted vertex number equals to $b(v)$ or there are no consumers left.
Finally, each pointer of the ad vertex $v$ points to the consumer vertex whose \emph{edge weight} is defined as the \textbf{threshold} of the weights of all neighbors. We denote this weight threshold as $\text{w}_\text{thr}(v)$. At the end of iteration, the neighbors whose edge weights are greater than $\text{w}_\text{thr}(v)$ are poured out by each vertex $v$. In this example, $\text{w}_\text{thr}(a)=2$ and $\text{w}_\text{thr}(b)=3$. Based on the analysis, the bottlenecks of the parallel greedy approaches and the way to alleviate them can be summarized as:
\begin{enumerate}[(1)]
\vskip -0.1cm
\setlength{\itemsep}{0.5pt}
\setlength{\parsep}{0.5pt}
\setlength{\parskip}{0.5pt}
\item The time complexity of the entire neighbor sorting process at step 1 is O($\delta(v)\text{log}\delta(v)$), where $\delta(v)$ is the degree of vertex $v$. If we know  $\text{w}_\text{thr}(v)$ for each advertiser beforehand, the sorting process of neighbors could be omitted. The reason is that we could consider $\text{w}_\text{thr}(v)$ a \textbf{pivot} (similar to the pivot in QuickSort) and only have to pour out the neighbors whose edge weights are greater than $\text{w}_\text{thr}(v)$, whose time complexity is thus reduced to O($\delta(v)$). Since $\delta(v)$ is in million scale in our case, the amount of time reduction is significant.
\item The existing parallel greedy approaches still needs hundreds of iterations before getting the solution for large-scale problems. The reason is that each ad vertex does not know how many neighbors should be poured out beforehand. Thus it has to iteratively move its pointer until finding the right one. However, if we know $\text{w}_\text{thr}(v)$ beforehand, only one iteration is needed to produce the solution since we could pour out all neighbors whose edge weights are greater than $\text{w}_\text{thr}(v)$) once.
\end{enumerate}

Once we know $\text{w}_\text{thr}(v)$ for each advertiser vertex beforehand, the time cost will be greatly reduced.
In next section, we present our approach \texttt{NeuSearcher}, which can make accurate predictions of $\text{w}_\text{thr}(v)$ for new problem instances based on the historical data and compute the matching solution in a faster manner based on the estimated $\text{w}_\text{thr}(v)$.

\section{Neural Searcher Framework}
\label{method}
\begin{figure}[htbp]
  \centering
  \includegraphics[width=3.3in]{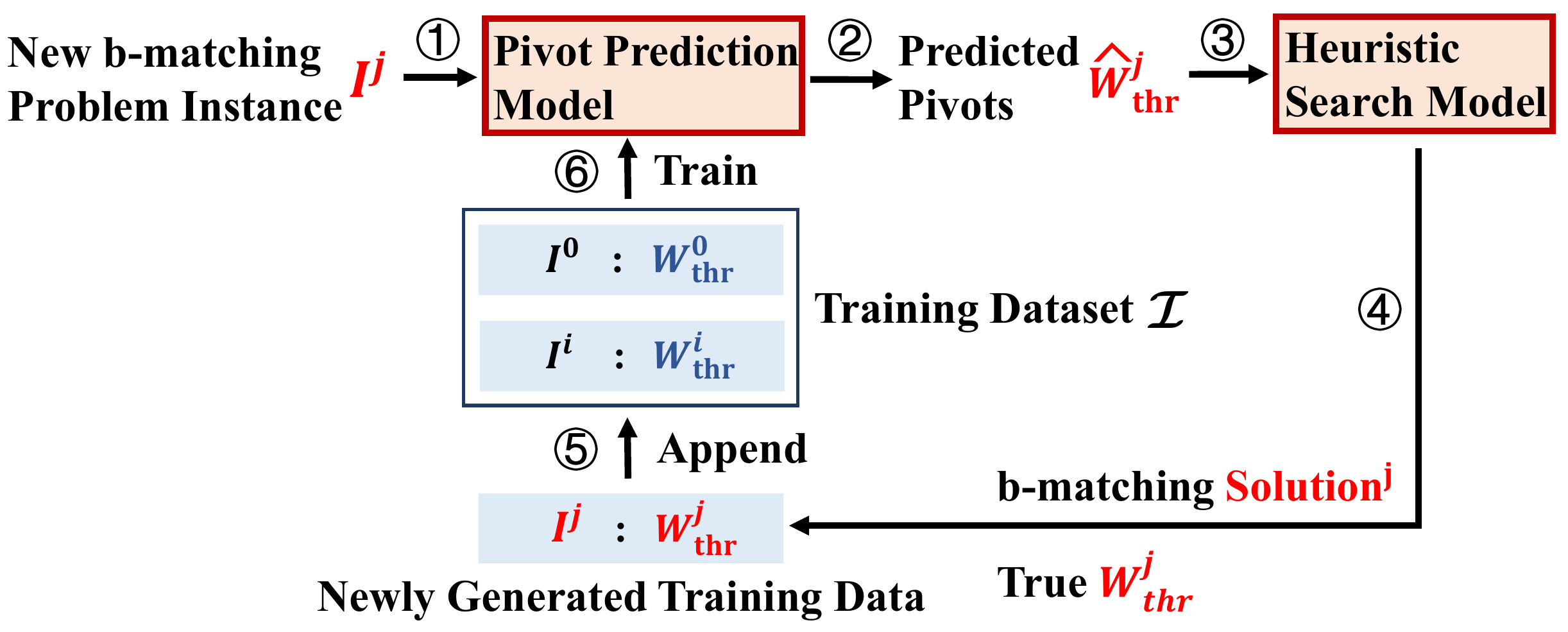}
  \caption{Neural Searcher Framework.}
  \label{Figure:NeuSearch}
\end{figure}

The proposed \texttt{NeuSearcher} is illustrated in Figure \ref{Figure:NeuSearch}, which consists of two phases. 
(1) \textbf{Offline training}: Given a set of already solved problem instances $\mathcal{I}=\{(I^i, W_\text{thr}^{i})\}$, where $I^i$ is a solved b-matching instance and $W_\text{thr}^{i}=\{\text{w}_\text{thr}(a), \forall a \in \mathbb{A}\}$ is a vector label containing a set of true weight threshold $\text{w}_\text{thr}(a)$ for all advertisers.
We train a predictive model to learn the mapping from each $I^i$ to $W_\text{thr}^{i}$. Specifically, a multichannel graph neural network is designed to make more accurate predictions.
(2) \textbf{Online solution computing}: Given a new problem instance $I^j$, we utilize the already trained model to quickly predicts $\text{w}_\text{thr}(v)$ for each ad $v$, denoted as $\hat{W}_\text{thr}^{j}$. Then, each predicted $\text{w}_\text{thr}(v)$ will be considered as a pivot value, which partitions the search space of the solution into 2 subsets. A better initial match solution could be quickly acquired within the subset with heavier edges. If all $\text{w}_\text{thr}(v)$ are correctly predicted, the initial solution is exactly the finally converged one. However, considering $\text{w}_\text{thr}(v)$ may have some deviation from the true value, we further design a parallel heuristic search model, which takes the coarse solution as input and efficiently fine-tunes it towards better qualities until convergence. Finally, we acquire the b-matching solution and the true $W_\text{thr}^{j}$. $(I^j,  W_\text{thr}^{j})$ is updated to $\mathcal{I}$, which will be further used to update the parameters of the predictive model.
\subsection{Pivot Prediction Model}
\label{GCN}
Given a graph with node features $X_v$, the target is to predict $\text{w}_\text{thr}(v)$ for each ad $v$. To build such a predictive model, the following factors should be take into consideration: (1) Since the b-matching is naturally defined in a graph, the designed model should have the capacity to capture the inherent structure (vertices, edges, constraints and their relationships) of the b-matching instances. (2) The model should be applicable to different size of graph instances and be capable of handling input dimension changes (different vertex has different number of neighbors). (3) In targeted advertising, the bipartite graphs are extremely unbalanced, i.e., $|\mathbb{C}| \gg |\mathbb{A}|$, which means the number of consumers (billions) is much larger than the number of advertisers (thousands). These considerations pose challenges to structural design of the model.
In this paper, we leverage Graph Neural Networks (GNNs) \cite{wu2019comprehensive} because they could maintain the graph structure and are well-defined no matter the input graph size and the input dimension. Modern GNNs follow a neighborhood aggregation strategy, where it iteratively updates the representation of a node by aggregating representations of its neighbors. However, since the bipartite graphs are unbalanced, i.e., $|\mathbb{C}|\!\gg\!|\mathbb{A}|$, simply applying GNN with a single-channel aggregate function (e.g., even the powerful sum-pooling \cite{xu2018powerful}) would result in significant loss of information.
\begin{figure}[t]
  \centering
  \includegraphics[width=2in]{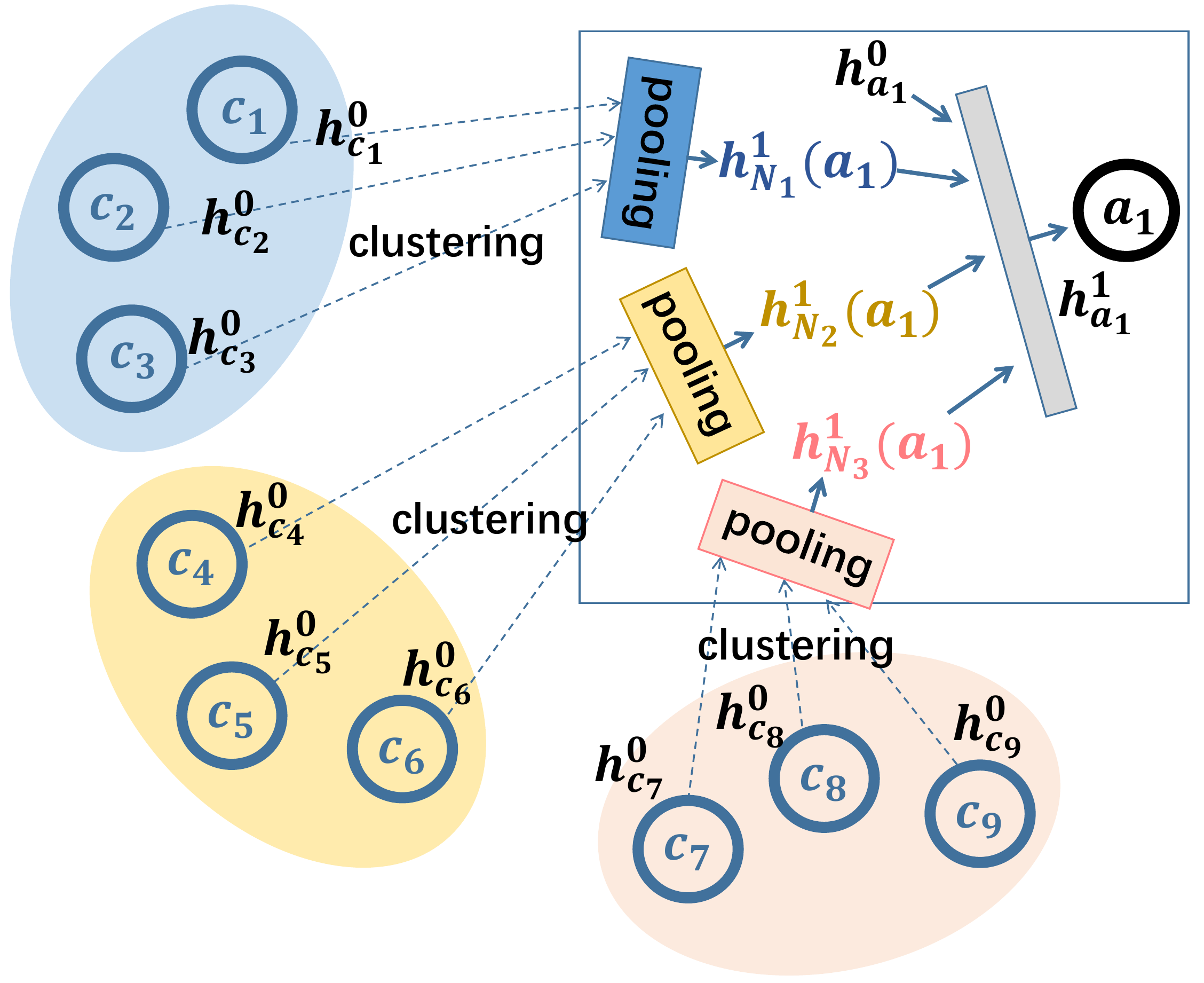}
  \caption{Illustration of our multichannel convolutional layer.}
  \label{Figure:gcn structure}
\end{figure}

To address this issue, we design a multichannel GNN which preserves more information during aggregating and improves its representational capacity. As in Figure \ref{Figure:gcn structure} (from an ad's view), we learn a \emph{differentiable} soft cluster assignment matrix for nodes at each layer of a GNN, mapping candidate nodes to a set of channels. Since the learned clustering procedure assigns different nodes to different channels while putting similar nodes together, we can naturally aggregate the nodes within the same channel through sum-pooling (since they are similar) while keeping all information among different channels using concat operation (since they are different). Thus, we obtain a distribution-style summarization of the neighbors' information.

We denote the learned cluster assignment matrix at layer $k$ as $S^{(k)}$$\in\!\mathbb{R}^{n_{a_i}\!\times\!c_k}$, where $c_k$ is the number of channels, $n_{a_i}$ is the number of neighboring consumers for advertiser $a_i$. Each row of $S^{(k)}$ corresponds to one of the $n_{a_i}$ neighboring consumers, and each column corresponds to one of the $c_k$ channels. Intuitively, $S^{(k)}$ provides a soft assignment of each neighboring consumer to a number of channels. Following the \emph{aggregate} and \emph{combine} paradigm \cite{wu2019comprehensive}, Equation \ref{Equation:mask} takes the neighbor embeddings $H_{n_{a_i}}^{k-1}$$\in\!\mathbb{R}^{n_{a_i}\!\times\!d_{k\!-\!1}}$ and aggregates them according to the cluster assignments $S^{(k)}$, generating neighbors' multichannel representations $\tilde{h}_{n_{a_i}}^{k}$. Then, the multichannel representations are flattened and combined (Equation \ref{Equation:flatten}) with ad $i$'s embedding $h_{a_i}^{k-1}$ at layer $k$-$1$, where $\big\|$ is the concat operator.
\begin{equation}
\label{Equation:mask}
\begin{aligned}
 \tilde{h}_{n_{a_i}}^{k}\! =\! ({S^{(k)}}^{\mathsf{T}}\cdot H_{n_{a_i}}^{k\!-\!1}) \!\in\! \mathbb{R}^{c_k\!\times\! d_{k-1}} \quad \rhd\text{AGGREGATE} 
\end{aligned}
\end{equation}
\begin{equation}
\setlength{\abovedisplayskip}{-0.1cm}
\label{Equation:flatten}
\begin{aligned}
 h_{a_i}^k\! \leftarrow\! \text{MLP}\left( \left[h_{a_i}^{k\!-\!1} \big\| \text{flatten}(\tilde{h}_{n_{a_i}}^{k})\right] \right) \quad  \rhd\text{COMBINE}
\end{aligned}
\end{equation}
 To generate the assignment matrix $S_{a_i}^{k}$ for each layer $k$, we apply a multi-layer perception (MLP) to the input neighbor embeddings $H_{n_{a_i}}^{k-1}$ of ad $a_i$, followed by a softmax layer for classification purpose:
\begin{equation}
\label{Equation:softmax}
\begin{aligned}
 S^{(k)} = \text{softmax}(\text{MLP}(H_{n_{a_i}}^{k-1})) \in \mathbb{R}^{n_{a_i} \times c_k}
\end{aligned}
\end{equation}
where $c_k$ is the number of clusters.
After $K$ layer aggregations, we acquire the ad node embeddings $h_v^{K}$ and pass them through an MLP and finally produce a single dimension output to predict $\text{w}_\text{thr}(v)$ for each ad $v$.
\begin{equation}
\label{Equation:sigmoid}
\begin{aligned}
 \hat{\text{w}}_\text{thr}(v) = \text{MLP}(h_v^{K}), \forall v \in \mathbb{A}
\end{aligned}
\end{equation}

\noindent\textbf{Training Details.}
Taking the already solved instances $\mathcal{I}=\{(I^i, W_\text{thr}^{i})\}$ as training data, we train the pivot prediction model end-to-end in a supervised fashion, using a mean-square error (MSE) loss. At the very beginning, when $\mathcal{I}$ is an empty set, we run the b-suitor over the recent problem instances to get the corresponding labels $W_\text{thr}^{i}$.

\subsection{Heuristic Search Model}
\label{model:inference}
During the online solution computing phase when given a new b-matching problem instance, we first call the pivot prediction model trained before to predict the pivot value (i.e., the weight threshold) $\text{w}_\text{thr}(v)$ for each ad vertex $v$. Further, to ensure the solution quality, we propose a parallel heuristic search algorithm as follows.
\begin{algorithm}[h]
    \caption{Parallel heuristic search algorithm}
    \label{Algorithm:finetunning}
    \begin{algorithmic}[1]
        \STATE \textbf{Input:} Bipartite graph $G=(\mathbb{C}, \mathbb{A}, \mathbb{E})$ and a constraint function $b(v), \forall v \in \mathbb{C} \cup \mathbb{A}$, an estimated $\text{w}_\text{thr}(a)$, $\forall a \in \mathbb{A}$. Each $c \in \mathbb{C}$ initializes a min-heap of size $b(c)$;
        \STATE \textbf{Output:} b-matching solution;
        \FOR{each vertex $a \in \mathbb{A}$ in parallel}
        \STATE Takes $\text{w}_\text{thr}(a)$ as the pivot and partitions the search space of all neighbors into 2 subsets;
        \STATE The heavier edges than the pivot are poured out; These edges are put into corresponding min-heaps;
        \STATE Count the number of currently reserved edges in the min-heaps. The number denoted as $\hat{b}(a)$;
        \ENDFOR
        \FOR{each iteration}
        \FOR{each vertex $a \in \mathbb{A}$ in parallel}
        \STATE Acquires $\hat{b}(a)$, the number of reserved edge in the min-heaps currently; Denotes $b_\delta(a)=\hat{b}(a)-b(a)$;
        \IF {$b_\delta(a)>0$}
        \STATE Recalls back $b_\delta(a)$ smallest edges preserved in the min-heaps and puts the ad vertices squeezed out by these $b_\delta(a)$ edges back into the min-heaps.
        \ELSIF {$b_\delta(a)<0$}
        \STATE Pours out another $|b_\delta(a)|$ neighbors in the descending order from the remaining neighbors.
        \ENDIF
        \ENDFOR
        \IF {$b_\delta(a)==0$ or no neighbors left, $\forall a \in \mathbb{A}$}
        \STATE return edges in all min-heaps as solution;
        \ENDIF
        \ENDFOR
    \end{algorithmic}
\end{algorithm}
The algorithm takes the estimated pivot value $\text{w}_\text{thr}(v)$ as input and quickly produce a initial solution (line 4-7). Then to ensure that the b-matching solution is exactly the same with the state-of-the-art greedy approaches, a fine-tuning process (line 8-20) is developed following the idea of the parallel b-suitor algorithm. The proof is presented as follow.

\noindent\textbf{The proof sketch.} In \cite{khan2016efficient} 3.2\&3.3, it proves that b-suitor achieves the same solution as serial greedy algorithm and the b-suitor finds the solution irrespective of the order of the vertices and the edges processed. Here we show that the high quality initial solution given by our method can be seen as an intermediate solution following some b-suitor processing order of vertices and edges. And since the rest fine-tuning process of Algo.1 is the same as b-suitor, our method naturally achieves exactly the same solution. Here we give the reason that our method can be seen as an intermediate solution. In Algo. 1, after the first pass of line 9-16, the solution given by our method (denoted as $S$) satisfies all constraints. In $S$, we define a set $P$ containing all poured out edges (including all reserved and squeezed out edges). The $S$ can be seen as an intermediate solution of b-suitor by selecting edges in $P$ following the descending weight order from an empty solution. This completes the proof.

\section{Experiments}
\subsection{Experimental Setup}
\noindent\textbf{Baselines.} We evaluate the performance of \texttt{NeuSearcher} against the following state-of-the-art baselines. (1) \emph{\textbf{optimal}}: We use Gurobi optimizer \cite{optimization2014inc} with an MIP formulation to compute the optimal solutions. (2) \emph{\textbf{serial greedy}}: The \emph{greedy} algorithm is a practical approximate algorithm which guarantees a 1/2 approximation ratio in the worst case \cite{avis1983survey,preis1999linear}. But in practical problems, its solutions are usually within 5\% percent of the optimal ones \cite{hougardy2009linear}. (3) \emph{\textbf{greedyMR}}: \cite{de2011social} adapt the serial greedy algorithm to the MapReduce environment. And greedyMR is one of the fastest parallel algorithms in computing b-matching problems. (4) \emph{\textbf{b-suitor}}: \emph{b-suitor} is the fastest (state-of-the-art) parallel approach for b-matching proposed by \cite{khan2016efficient}. All experiments are conducted on an Intel(R) Xeon(R) E5-2682 v4 processor based system with a memory of 128G. All codes were developed using C++ 11 multi-thread.

\noindent\textbf{Datasets.} We evaluate \texttt{NeuSearcher} on both open and industrial datasets. Table \ref{tab:datasets} summarizes the dataset properties. Each of the first 7 datasets (adv \#1 to \#7) has more than a billion edges, which are collected from the e-commerce platform of Alibaba for seven consecutive days.
Due to the memory limit (128G), we cannot calculate the exact solution using Gurobi optimizer for the first 7 datasets. Thus, we compare the matching quality of the approximate algorithms relative to the exact solution on the other 3 open datasets (Amazon review data \cite{he2016ups} and MovieLens data \cite{harper2016movielens}).

\begin{table}[htbp]
  \scalebox{0.94}{
  \begin{tabular}{lcccc}
    \toprule
    Graph & \# $\mathbb{C}$ & \# $\mathbb{A}$ & \# $\mathbb{E}$ & Avg. Deg. of $\mathbb{A}$ \\
    \midrule
     adv \#1 to \#7 & 236M  & 46k & 1B & 24k \\
    \midrule
    MovieLens10M    &   69k     &    10k  & 10M & 936.6\\
    MovieLens20M    &  138k     &    26k  & 20M & 747.8\\
    RatingsBooks   & 8M   & 2M  & 22M &  9.7\\
    \bottomrule
  \end{tabular}
  }
  \caption{The structural properties of the datasets.}
  \label{tab:datasets}
\end{table}

\noindent\textbf{Other Settings.} For the 7 advertising datasets, we use the first 4 for training, the 5th for validation and the last 2 for testing. For the other 3 open datasets, we add Gaussian noise with mean 0.0 and variance 0.1 to the edge weights and generate 4 more datasets for each (3 for training and 1 for validation).
In following experiments, unless otherwise mentioned, we fix $b(v)=0.5*\delta(v), \forall v \in \mathbb{A}$ and set $b(v)=min\{b,\delta(v)\}$, $\forall v \in \mathbb{C}$, where $\delta(v)$ is the degree of $v$ and $b\!=\!\text{avg}\{\delta(v), \forall v \in \mathbb{C}\}$. For hyperparameters, we set $K\!=\!2$, $c_k\!=\!16$ after grid-search optimization.

\begin{table}[!htbp]
  \scalebox{0.94}{
  \begin{tabular}{llll}
  \toprule
  \multirow{4}{*}{Graph} & serial greedy &                          &    \\
                       & greedyMR      & \multirow{1}{*}{optimal} & \multirow{1}{*}{Quality}         \\
                       & b-suitor      & \multirow{1}{*}{(Gurobi)} & \multirow{1}{*}{in \%}  \\
                       & NeuSearcher   &                          &                                 \\
  \midrule
  MovieLens10M           & 29,995,076.5   & \textbf{30,510,066}      & 99.05                                        \\
  MovieLens20M           & 60,247,629.5   & \textbf{61,194,930}      & 98.45                                        \\
  RatingsBooks           & 77,213,078    & \textbf{79,068,583}      & 97.65                                        \\
  \midrule
  adv \#6                & 28,724,740.17 & \multicolumn{2}{c}{out-of-memory error}                       \\
  adv \#7                & 28,150,245.37 & \multicolumn{2}{c}{out-of-memory error}                        \\
  \bottomrule
  \end{tabular}
  }
  \caption{The solution quality comparison (best in bold).}
  \label{tab:matching-weights}
\end{table}

\subsection{Solution Quality Comparison}

We compare the matching value of the optimal solution as well as all approximate baselines with our \texttt{NeuSearcher} in Table \ref{tab:matching-weights}. Among the experimental results over all 5 datasets, the 4 approximation approaches, i.e., serial greedy, gredyMR, b-suitor and our NeuSearcher all find exactly the same set of matched edges with the same matching values. We summarize their results in the same column due to space limitation. Besides, in Table \ref{tab:matching-weights}, we see that although the approximate approaches theoretically can only guarantee 1/2 approximation in the worst case, they find more than 97\% of the optimal weight for the 3 open datasets. The highest approximation ratio of the approximate approaches achieved is 99.0\%. For problems with larger sizes, the Gurobi fails to compute an optimal solution due to the memory limit (128G). This indicates that faster approximate approaches are good alternatives in solving large-scale b-matching problems and our \texttt{NeuSearcher} achieves the state-of-the-art solution quality.

\begin{table*}[!htbp]
\setlength{\belowcaptionskip}{-0.2cm}
  \begin{tabular}{lcccccc}
    \toprule
    Graph            & serial greedy & greedyMR & b-suitor & NeuSearcher (multichannel GNN)  & optimal (Gurobi) \\
    \midrule
    MovieLens 10M    &   92.952      &  32.705  & 35.889   &   \textbf{15.141}            &     742.795               \\
    MovieLens 20M    &  190.221      &  91.462  & 78.588   &   \textbf{35.059}            &    2355.614               \\
    Ratings\_Books   &  235.607      &  53.387  & 34.627   &   \textbf{14.212}            &   44376.265 (12.3 hour)    \\
    \midrule
    adv \#6  & 15352.075      & 1875.154  & 410.270  &  \textbf{199.423}           & out-of-memory error\\
    adv \#7  & 14831.704      & 1893.876  & 426.359  &  \textbf{201.094}           & out-of-memory error\\
    \bottomrule
  \end{tabular}
  \caption{The runtimes (in seconds) of b-matching computation, where lower values are better (best in bold).} 
  \label{tab:time-comparison}
\end{table*}

\subsection{Runtime Comparison}
We provide the online solution computing time of our approach as well as runtimes of other methods over 5 datasets in Table \ref{tab:time-comparison}.
We use the same evaluation metric (clock time) to record the computing time. All results are averaged over 10 rounds. For all approaches, only CPUs are used for the sake of
fair comparison, though our model can be accelerated using GPUs. 
In Table \ref{tab:time-comparison}, we see that even for the smaller open datasets, Gurobi still needs hours to compute the solutions, which is intolerable. For larger datasets adv \#6 and \#7, Gurobi fails and causes out-of-memory error.
On the contrary, all approximate approaches are much faster than the exact algorithm. Our \texttt{NeuSearcher} with the designed multichannel GNN computes the same solutions at the fastest speed by reducing more than 50\% computing time. 
Among other approximate methods, \emph{b-suitor} runs faster than \emph{greedyMR} and requires less iterations to compute the results. The serial \emph{greedy} algorithm is the slowest since it requires a global sorting and a sequential decision process.
Combining Table \ref{tab:matching-weights} with \ref{tab:time-comparison}, we conclude that our \texttt{NeuSearcher}
can achieve a much faster speed, while still acquire exactly the same matching solution with the state-of-the-art approaches.

\subsection{Convergence Analysis}
\begin{figure}[htbp]
\setlength{\abovecaptionskip}{0.2cm}
  \centering
  \includegraphics[width=2.6in]{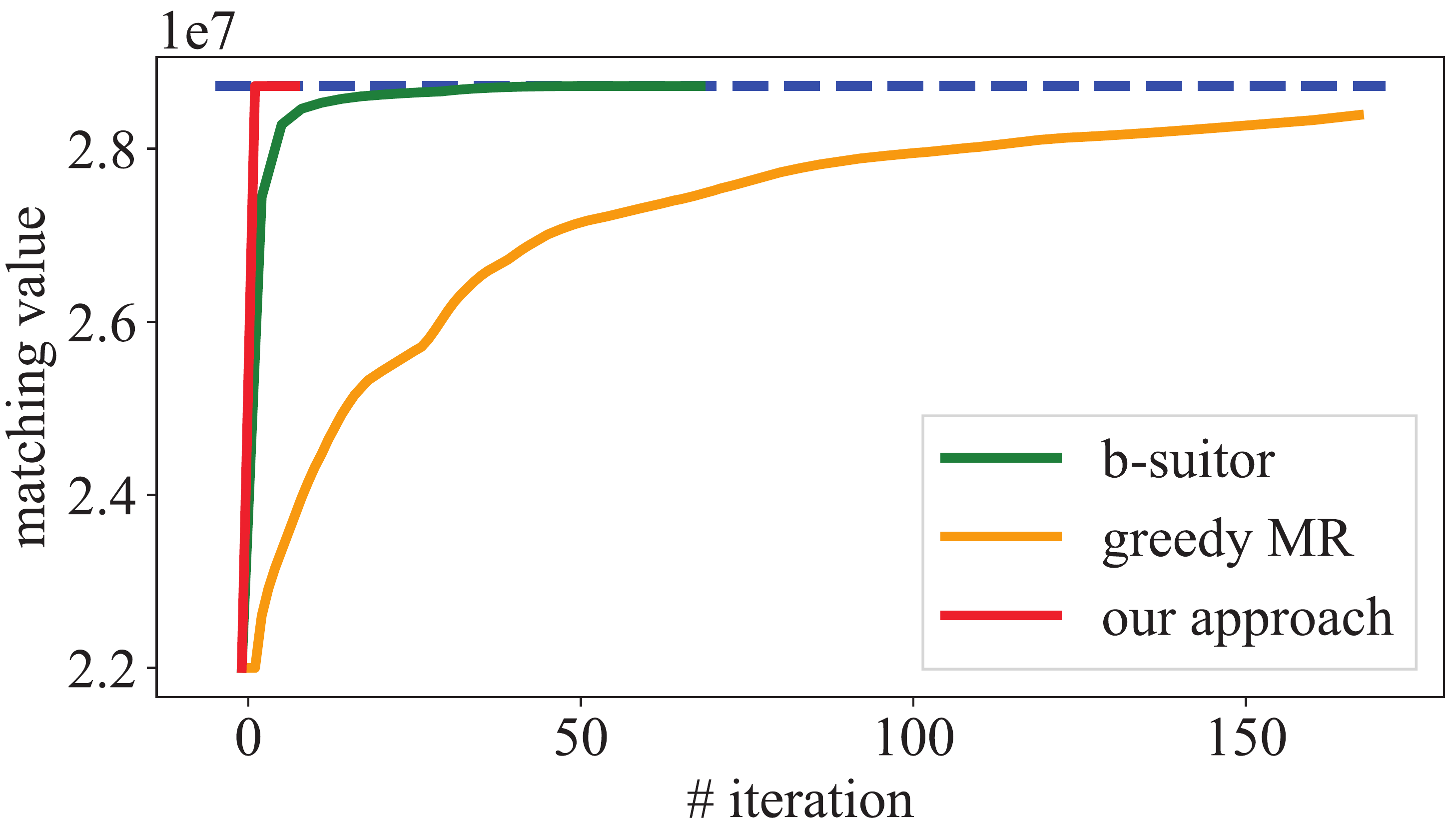}
  \caption{Matching value (of adv \#6) by the greedyMR, b-suitor and our \texttt{NeuSearcher} as a function of the number of iterations.}
  \label{Figure:iteration and time}
\end{figure}

To better analyze the computing process of the three parallel approximate algorithms: greedyMR, b-suitor and our \texttt{NeuSearcher}, we plot their solution computing process in Figure \ref{Figure:iteration and time} using adv \#6 dataset as an example. We see that our approach requires the fewest (15) iterations to compute the solution. However, the b-suitor needs 68 iterations and the greedyMR needs 358 iterations.
The reason is that the neural net captures the correlations between the problem structure and the weight threshold $\text{w}_\text{thr}(v)$ (pivot), which significantly reduces the search region of the solution space. Then, the following heuristic search algorithm benefits more from a better jumping start and only needs few steps to fine-tune the initial solution towards convergence.

\subsection{Ablation Study: Effect of multichannel GNN}
\begin{figure}[!htbp]
\setlength{\abovecaptionskip}{0.15cm}
\setlength{\belowcaptionskip}{-0.2cm}
\centering
\subfigure[Training loss] {\includegraphics[height=1.09in,width=1.6in,angle=0]{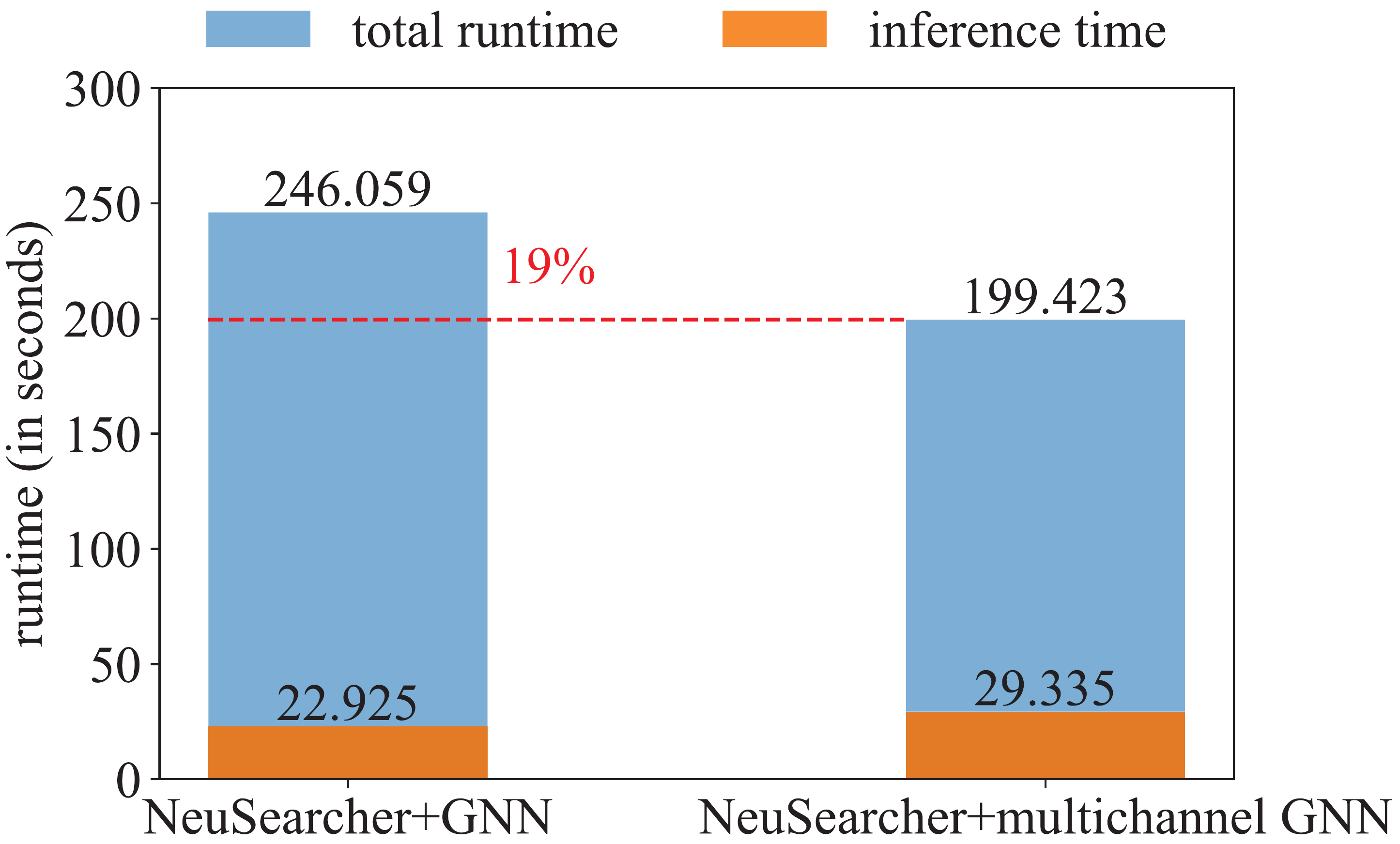}}
\subfigure[Validation loss] {\includegraphics[height=1.1in,width=1.65in,angle=0]{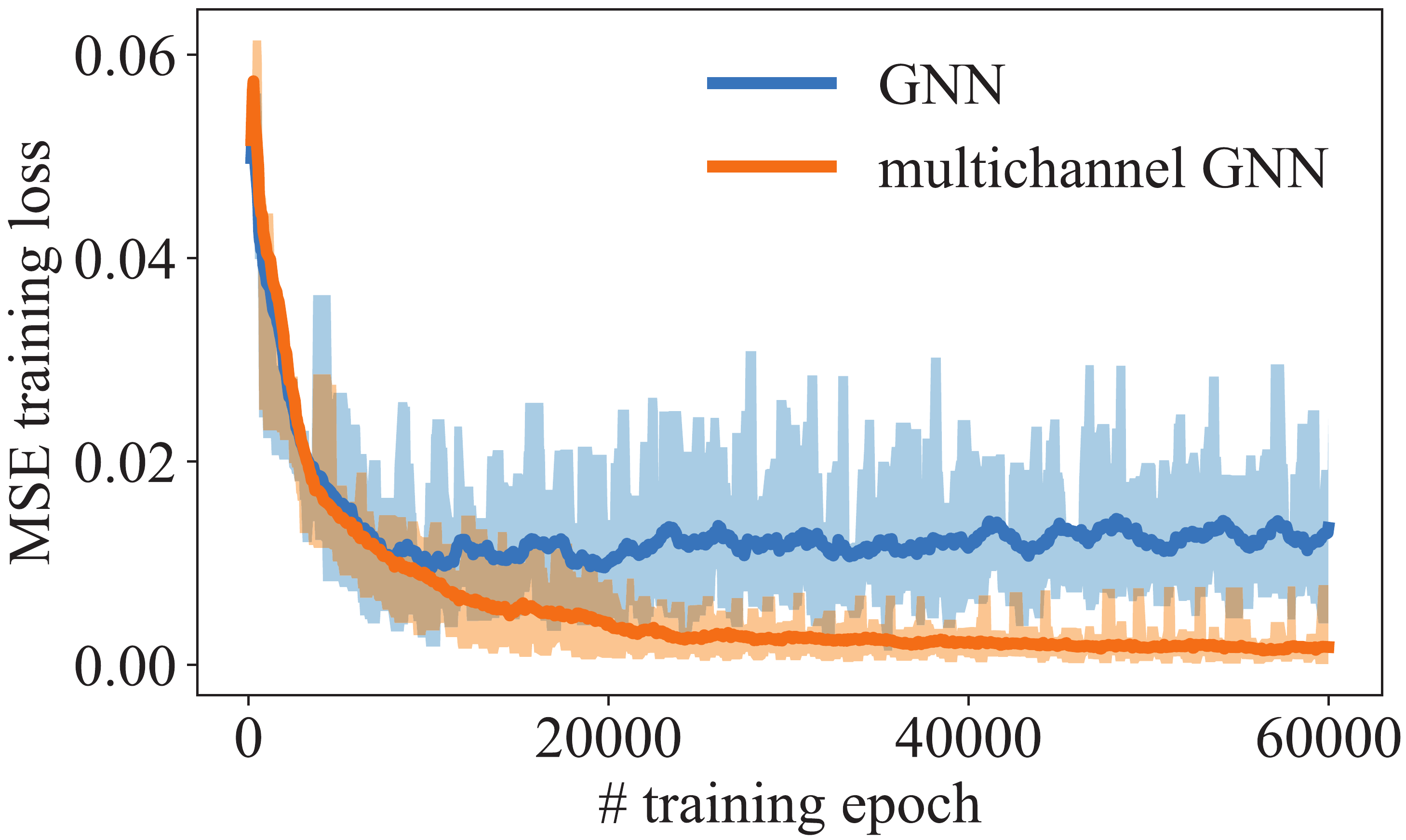}}
\caption{Comparison of the runtime and validation loss of multichannel GNN and GNN in adv \#6 dataset.}
\label{Figure:runtime and loss}
\end{figure}

In Figure \ref{Figure:runtime and loss} (a), we compare the detailed solution computing time of NeuSearcher with multichannel GNN and NeuSearcher with GNN. We see NeuSearcher with multichannel GNN is the fastest, which reduces 19\% overall computing time. Besides, we also separately compare the two inner stages of the solution computing: 1) pivot prediction (inference) and 2) fine-tuning. We see though the inference time of multichannel GNN is slightly longer than GNN, the overall time cost is much smaller, which indicates multichannel GNN provides a more precise pivot value by which reducing the subsequent fine-tuning steps.
In detail, NeuSearcher with multichannel GNN only needs 15 fine-tuning iterations while NeuSearcher with GNN needs 29 iterations. Similar evidences can also be found in Figure \ref{Figure:runtime and loss} (b), where we compare the validation losses of the two models. For the reason that the multichannel GNN has a better representational ability and generalizes well, the validation loss is much lower.

\section{Conclusion}
To the best of our knowledge, we are the first to integrate deep learning methods to accelerate solving practical large-scale b-matching problems.
Our \texttt{NeuSearcher} transfers knowledge learned from previous solved instances to save more than 50\% of the computing time. We also design a parallel heuristic search algorithm to ensure the solution quality exactly the same with the state-of-the-art approximation algorithms.
Given highly unbalanced feature of the advertising problem, we design a multichannel graph neural network to encode the billions consumers and their diverse interests to improve the representation capability and accuracy of the pivot prediction model. Experiments on open and real-world large-scale datasets show \texttt{NeuSearcher} can compute nearly optimal solution much faster than state-of-the-art methods.

\section*{Acknowledgments}
The work is supported by the Alibaba Group through Alibaba Innovative Research Program, the National Natural Science Foundation of China (Grant Nos.: 61702362, U1836214) and the new Generation of Artificial Intelligence Science and Technology Major Project of Tianjin under grant: 19ZXZNGX00010.

\bibliographystyle{named}
\bibliography{reference}

\end{document}